\documentclass[a4paper]{jpconf}
\usepackage{graphicx}
\usepackage{iopams}
\begin{document}
\title{Cosmological constraints on heavy sterile neutrinos}

\author{L Mastrototaro}

\address{Dipartimento di Fisica ``E.R Caianiello'', Università degli Studi di Salerno, Via Giovanni Paolo II, 132 - 84084 Fisciano (SA), Italy.}
\address{Istituto Nazionale di Fisica Nucleare - Gruppo Collegato di Salerno - Sezione di Napoli, Via Giovanni Paolo II, 132 - 84084 Fisciano (SA), Italy.}

\ead{lmastrototaro@unisa.it}

\begin{abstract}
We consider heavy sterile neutrinos $\nu_s$,  with mass in the range 10 MeV $\lesssim m_s  \lesssim  m_{\pi} \sim 135$ MeV, thermally produced in the early universe and freezing out after the QCD phase transition. The existence of these neutrinos may alter the value of the effective number of neutrino species $N_{\rm eff}$, measured by the cosmic microwave background (CMB), and the ${}^4$He production during the BBN. We provide a detailed account of the solution of the relevant Boltzmann equations and we identify the parameter space constrained by current 
Planck satellite data and forecast the parameter space probed by future Stage-4 ground-based CMB (CMB-S4) observations.
\end{abstract}

\section{Introduction}

Sterile neutrino $\nu_s$ emerges naturally in almost all the extensions of the Standard Model, like the Neutrino Minimal Standard Model ($\nu$MSM)~\cite{Asaka:2005an}, where such particle can be related to fundamental problems of particle physics as the origin of neutrino mass, the baryon asymmetry in the early universe and the nature of dark matter. \\
Constraints on heavy sterile neutrinos can arise from cosmological arguments: $\nu_s$, with mass in the range 10 MeV $\lesssim m_s  \lesssim  m_{\pi} \sim 135$ MeV, can be produced in the early universe and then decay into lighter species. If they mostly decay after the active neutrinos decoupling, the effective number of neutrino species $N_{\rm eff}$ is modified and the non-thermal $\nu_e$ and $\bar{\nu}_e$ populations ,entering in the weak interactions, alter also the \emph{neutron-to-proton ratio} which rules the abundance of the primordial yields \cite{Dolgov:2000pj,Dolgov:2000jw,Ruchayskiy:2012si,Sabti:2020yrt}. This effect is particularly relevant for $^4$He which can be taken into account in the primordial Helium mass fraction parameter $Y_p$.\\
Our work aims to perform a calculation without approximation of $\nu_s$ evolution and decoupling in the early universe, computing the changes on  $N_{\mathrm{eff}}$ and  $Y_p$ and find the constraints on the sterile neutrino parameter space.
In Sec.~\ref{evolution} we discuss and solve the kinetic equations describing the sterile neutrino evolution in the early universe.
In Sec.~\ref{constraints} we characterise the impact of heavy sterile neutrino decays on active neutrinos and we present the current constraints and forecasts on sterile neutrino parameter space from BBN and CMB data.
Finally, in Sec.~\ref{summary} we summarise our results and conclude.\\
The results presented here are based on our work Ref.~\cite{Mastrototaro:2021wzl} to which we address the interested readers for further details.

\section{Sterile and active neutrino evolution in the early universe}
\label{evolution}
\subsection{Equations of evolution}
We consider heavy sterile neutrinos mixed dominantly with one active neutrino 
$\nu_\alpha$ ($\alpha=e,\mu,\tau$):
%..........................
\begin{equation}
\nu_\alpha=\cos \theta_{\alpha s} \nu_\ell + \sin\theta_{\alpha s} \nu_H \quad\quad
\nu_s=-\sin \theta_{\alpha s} \nu_\ell + \cos\theta_{\alpha s} \nu_H \,\ ,
\end{equation}
%.........................
where $\nu_\ell$ and $\nu_H$ are a light and a heavy neutrino mass eigenstate, respectively. It is possible to relate the mixing angle to the unitary mixing matrix $\rm U$, where
%.....................
\begin{equation}
|\rm U_{\alpha s}|^2 \simeq \frac{1}{4} \sin^2 2 \theta_{\alpha s}\simeq \theta_{\alpha s}^2 \,\ .
\end{equation}
%........................
To treat the particle evolution in the early universe, it is useful to define the dimensionless variables
%...............
\begin{equation}
x= m a \,\ \,\ \,\ \,\ \,\ \,\  y=p a \,\ \,\ \,\ z=T a \,\ ,
\label{eq:eom}
\end{equation}
%.............
where $m=1~\mathrm{MeV}$ and the function $a$ can be normalized to obtain that $ z=1$ when all particles are in thermal equilibrium.
In terms of these variables, we can write the equations of evolution for the neutrinos distribution function $f_{\nu_a}$ as \cite{Dolgov:2000pj,Dolgov:2000jw}
%......................
\begin{equation}
H x \partial_x f_{\nu_\alpha} = I_{\nu_\alpha}[f_{\nu_\alpha}] \,\ \,\ \,\ \,\ \nu_\alpha=\nu_e,\nu_\mu, \nu_\tau, \nu_s .
\label{eq:active}
\end{equation}
%...................
In the previous equation, ${H}$ is the cosmic expansion Hubble rate and $I_{\nu_\alpha}[f_{\nu_\alpha}]$ is defined as
%%%%%%%%%%%%%%
\begin{equation}
I[f_\nu]=\frac{1}{2E}\int\prod_i\left(\frac{d^3p_i}{2E_i(2\pi)^3}\right)\prod_f\left(\frac{d^3p_f}{2E_f(2\pi)^3}\right)  (2\pi)^4 \delta^{(4)}\left(\sum_i p_i-\sum_fp_f\right)|M_{fi}|^2F(f_i,f_f) \,\ , 
\label{general_coll_integrala}
\end{equation}
%%%%%%%%%%%%%%%%%%%%%%%
with $|M_{fi}|^2$ the sum of the squared-matrix elements over initial and final states, divided by the spin multiplicity of the state of interest and the statistical factor
%%%%%%%%%%%%%
\begin{equation}
F(f_i,f_f)=-\prod_if_i\prod_f(1- f_f)+\prod_i(1- f_i)\prod_ff_f \,\ ,
\label{Ffactor}
\end{equation}
%%%%%%%%%%%%%%%%%%%%%%
where $f_{i,f}$ are the distributions of the particles in the initial or final states respectively. In Eq.~(\ref{general_coll_integrala}), $I[f_{\nu_s}]$ contains decay and scattering processes,
listed in Table~I and II of Ref.~\cite{Mastrototaro:2021wzl} respectively, while  $I[f_{\nu_a}]$ contains scattering processes among active neutrinos and the sterile neutrino decay injection term, which is calculated referring to the processes in Table I in~\cite{Mastrototaro:2021wzl}. Moreover, we take into account oscillations among active neutrinos considering
%...................
\begin{equation}
I[f_\alpha] \to \sum_\beta P_{\alpha \beta} I[ f_\beta]\,,
\end{equation}
%...................
where $P_{\alpha \beta}$ are the time-averaged transition probabilities (see e.g.~\cite{Ruchayskiy:2012si,Sabti:2020yrt}).\\
Finally, to get the time evolution of sterile and active neutrino distributions, we have also to take into account the plasma temperature evolution with the conservation of the total energy density
%...........................
\begin{equation}
\frac{d}{dx}{\bar\rho}(x)=\frac{1}{x}({\bar \rho}-3 {\bar P}) \,\ ,
\label{eq:contin}
\end{equation}
%.............................
where ${\bar \rho}$ and ${\bar P}$ are the comoving energy density and pressure of the primordial plasma. From that, we can obtain the ``time-temperature'' evolution as shown in Ref.~\cite{Mastrototaro:2021wzl}.
%%%%%%%%%%%%%%%%
\subsection{Evolution of heavy sterile neutrinos}
%%%%%%%%%%%%%%%%

\begin{figure}
\centering
    \includegraphics[scale=0.5]{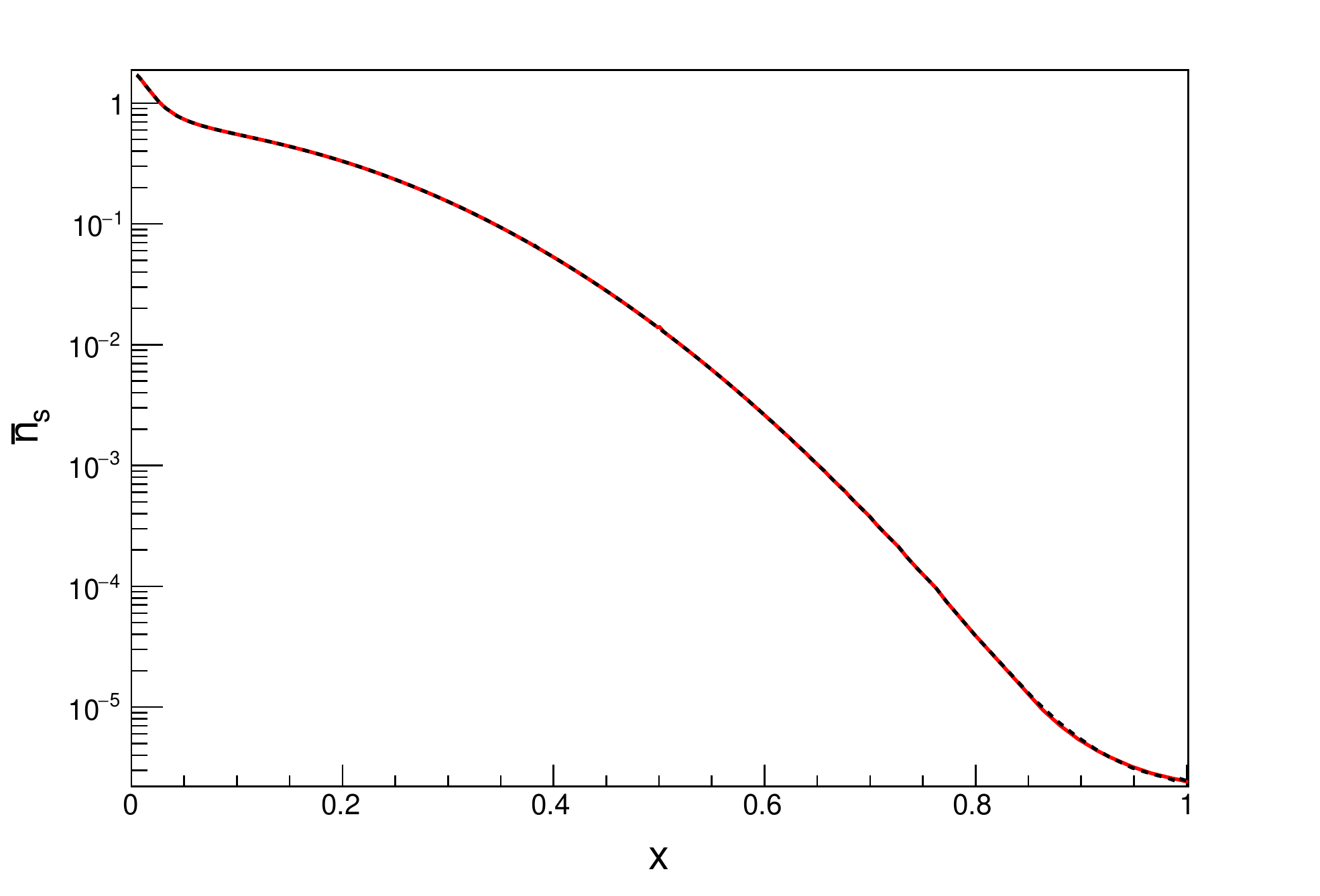}
    \caption{Comparison between our results (red, solid) and Ref.~\cite{Dolgov:2000jw} (black, dashed) on the comoving density of the sterile species for $m_s=100~\mathrm{MeV}$ and $\tau_s=0.055~\mathrm{s}$.}
    \label{ConfrontoDolgov}
\end{figure}

In~\cite{Dolgov:2000jw}, an analytical solution of Eq.~(\ref{eq:active}) was provided under the following assumptions:
%%%%%%%%%%%%
\begin{itemize}
\item[\emph{(i)}] The equilibrium distribution functions ``inside'' the collisional integral [Eq.~(\ref{general_coll_integrala})]  are taken in the Boltzmann approximation, neglecting the Pauli blocking factors.
\item[\emph{(ii)}] Electrons are considered massless.
\end{itemize}
%%%%%%%%%%
To test our code, we have first compared our results with that obtained in Ref.~\cite{Dolgov:2000jw}, finding excellent agreement as it is possible to see in Fig~\ref{ConfrontoDolgov} where we show the comoving density of the sterile specie. 
%%%%%%%%%%%%%%%%%%%%%%%%%%%%%%%%%%%%%%%%%%%%%%%%%%%%%%%%%%%%%%%%%%%%%%%%%%%%%%%%%%%%%%
\\
%%%%%%%%%%%%%%%%%%%%%%%%%%%%%%%%%%%%%%%%%%%%%%%%%%%%%%%%%%%%%%%%%%%%%%%%%%%%%%%%%%%%%%%
Then, we solve the sterile neutrino kinetic equations numerically, without approximations, for both sterile neutrinos and active neutrinos.  Following the well-known technique developed in~\cite{Hannestad:1995rs}, it is possible to  analytically reduce the nine-dimensional 
collisional integral into a three-dimensional one, which is then integrated numerically. We developed an equivalent technique for the decay processes. Our results are obtained by starting from an initial thermal equilibrium for all species, from a temperature $T=\mathrm{min}[2\,m_s,150\,{\rm MeV}]$. This approach is justified considering the obtained decoupling temperature listed in Table~\ref{TD} which are always below the QCD phase transition temperature $\sim 150$ MeV.
%%%%%%%%%%%%%%%%%%%%%

%...........................
\begin{table}
\caption{Sterile neutrino decoupling temperature $T_D^n$ obtained solving the Boltzmann equation numerically for some masses $m_s$ and mixing $|\rm U|^2_{\tau s}$ between the sterile neutrino and $\nu_\tau$. We have reported also the $\nu_s$ life time,$\tau$.}
\vspace{.5cm}
\centering
\begin{tabular}{cccc}
\hline
$m_s~[\mathrm{MeV}]$&  $|\rm U|^2_{\tau s}$& $\tau~[\mathrm{s}]$& $T^n_D~[\mathrm{MeV}]$\\
\hline
$20.0$& $2.6\times10^{-2}$& $3.0\times 10^{-1}$& $4.35$\\
$40.0$& $2.8\times10^{-3}$& $8.8\times 10^{-2}$& $9.24$\\
$60.0$& $5.5\times10^{-4}$& $6.0\times 10^{-2}$& $16.83$\\
$80.0$& $1.5\times10^{-4}$& $5.0\times 10^{-2}$& $26.53$\\
$100.0$& $5.8\times10^{-5}$& $4.4\times 10^{-2}$& $37.10$\\
$130.0$& $1.6\times10^{-5}$& $4.2\times 10^{-2}$& $59.13$\\
\hline
\end{tabular}
\label{TD}
\end{table}
%..................................
%%%%%%%%%%%%%%%%%%%%%%%%%%%%%%%%%%%%%%%%%%%%%%%%%%%
\section{Impact on cosmological observables}
%%%%%%%%%%%%%%%%%%%%%%
\label{constraints}
%%%%%%%%%%%%%%%%%%%%%%%%
After the distribution functions and temperature evolution are found, we relate them to the observables  $N_{\mathrm{eff}}$ (notably at the CMB epoch) and $Y_p$ (notably at the BBN epoch) to derive the constraints. The \emph{effective number of neutrinos}, $N_{\rm eff}$,  is defined as:
\begin{equation}
N_{\rm eff}(x)=  \frac{\rho_\gamma^{\rm inst}}{\rho_\gamma}\sum_{i\neq \rm e.m.}\frac{\rho_i}{\rho_{\nu_0}}= \left(\frac{z_0(x)}{z(x)} \right)^4 \left(3+ \frac{\Delta \rho_{\nu_e}}{\rho_{\nu_0}} + 
\frac{\Delta \rho_{\nu_\mu}}{\rho_{\nu_0}}+\frac{\Delta \rho_{\nu_\tau}}{\rho_{\nu_0}} + \frac{\rho_{\nu_s}}{\rho_{\nu_0}} \right) \,.
\label{eq:Neff}
\end{equation}
%................................
The other important parameter affected by the existence of a massive sterile neutrino, $Y_p$, is calculated in the Born approximation: a precise standard model calculation, $Y_{p, {\rm SM}}^{\rm prec}$ is obtained from \texttt{Parthenope}~\cite{Pisanti:2007hk,Consiglio:2017pot} and this result is rescaled via the ratio of the Born estimate of the $Y_p$ for the $\nu_s$ model, $Y_{p, \nu_s}^{\rm Born}$, over the Born standard model calculation $Y_{p, {\rm SM}}^{\rm Born}$, as 
\begin{equation}
Y_p=Y_{p, {\rm SM}}^{\rm prec}\, \frac{Y_{p, \nu_s}^{\rm Born}}{Y_{p, {\rm SM}}^{\rm Born}} \,\ .
\label{Yp}
\end{equation}
 Each term of the fraction in Eq.~(\ref{Yp}) can be estimated as shown in~\cite{Mastrototaro:2021wzl}.
%%%%%%%%%%%%%%%%%%%%%%%%%%
\subsection{Constraints and forecasts}\label{C&F}
%%%%%%%%%%%%%%%%

\begin{figure}
\centering
\includegraphics[scale=0.5]{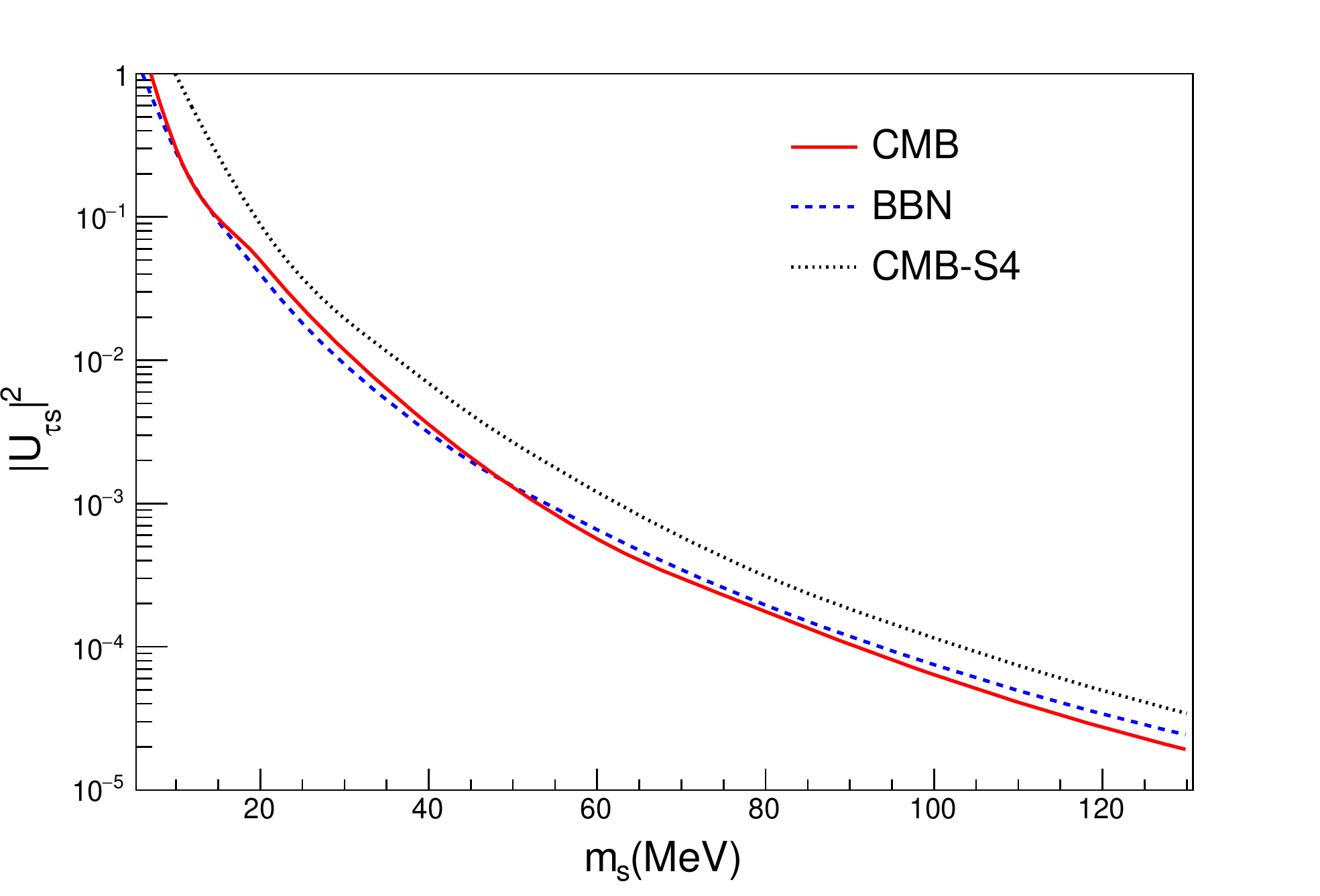}
\caption{Bounds in the plane $(m_s,|\mathrm{U}|^2_{\tau s})$ obtained from CMB (red curve) and BBN-$Y_p$ (blue curve), as well as forecast sensitivity of CMB-S4 (black curve), for a sterile neutrino mixed with  $\nu_\tau$ (or $\nu_\mu$). The $2\sigma$ excluded region is the one under the curves.}
\label{bound_t}
\end{figure}

\begin{figure}
\centering
\includegraphics[scale=0.5]{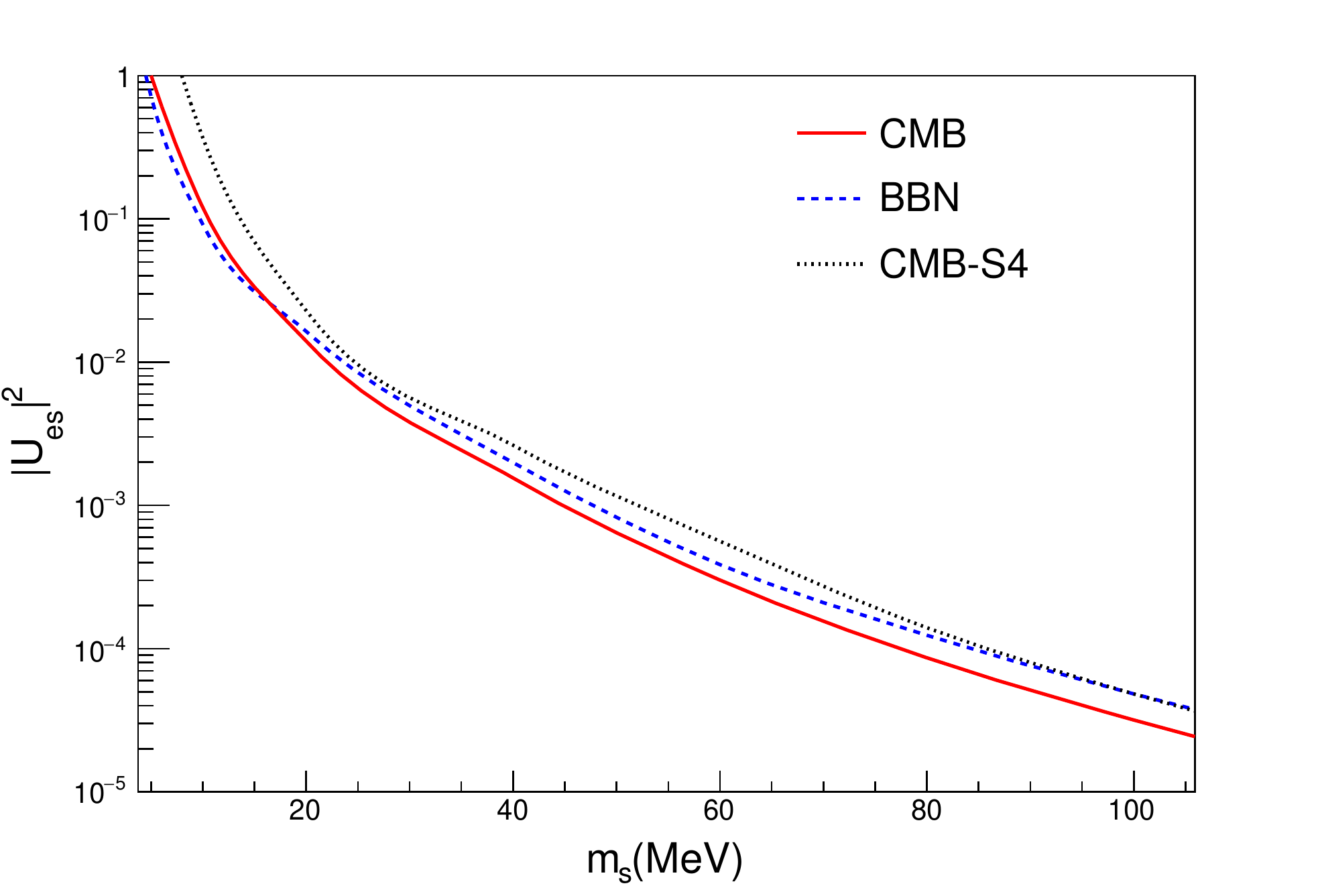}
\caption{Bounds in the plane $(m_s,|\rm U|^2_{es})$ obtained from CMB (red curve) and BBN-$Y_p$ (blue curve), as well as forecast sensitivity of CMB-S4 (black curve), for a sterile neutrino mixed with  $\nu_e$. The $2\sigma$ excluded region is the one under the curves.}
\label{bound_e}
\end{figure}

To obtain constraints on heavy sterile neutrinos, we compare our results on $N_{\rm eff}$ and $Y_p$ with both the latest CMB and BBN measurements. For BBN, we use the current bound $Y_p=0.245\pm 0.006$, while concerning CMB and CMB-S4 we use a reduced Gaussian likelihood matrix following~\cite{Sabti:2020yrt,Mastrototaro:2021wzl}. Our results from CMB measurements and from BBN at $2\sigma$ are shown in Fig.~\ref{bound_t} for the constraints on the mixing with $\nu_\tau$ ($|\mathrm{U}|^2_{\tau s}$) and in Fig.~\ref{bound_e} the corresponding results for the mixing with $\nu_e$ ($|\mathrm{U}|^2_{e s}$). We obtain that the CMB provides already the best constraints for  $m_s\lesssim \mathcal{O}(20~\mathrm{MeV})$, while BBN takes over at larger masses. However, we expect that the future CMB-S4 experiments will lead to the strongest constrain in the whole range of parameter space considered here if performing close to expectations. 
%%%%%%%%%
%%%%%%%
%%%%%%%%%
%%%%%%%%%%%%%%%%%%%%%%%%%%%%%%%%%%%%%%%
\section{Conclusions}\label{summary}
We have solved the exact Boltzmann equation for the evolution of active and sterile neutrino, with $10~\mathrm{MeV}\lesssim m_s \lesssim 135~\mathrm{MeV}$ while taking into account the temperature evolution of electrons and photons. At the end of the evolution, we have compared the values of $N_{\mathrm{eff}}$ and $Y_p$ with the current constraints from Planck. While currently, CMB is more constraining at low masses and BBN dominates at high masses, we expect the future CMB-S4 experiments will constraint even more the $\nu_s$ parameter space unless the systematic error affecting the astrophysical determinations of $Y_p$ can be significantly reduced.
%%%%%%%%%%%%%%%%%%%%%%%%%%%%%%%%%%%%%%%%

\ack
L.M. is grateful to Pasquale Dario Serpico, Alessandro Mirizzi and Ninetta Saviano who collaborated in the realization of the work upon which this contribution is based.
The work of L.M. is supported by the Italian Istituto Nazionale di Fisica Nucleare (INFN) through the ``QGSKY'' project and by Ministero dell'Istruzione, Universit\`a e Ricerca (MIUR).

\section*{References}
\bibliography{iopart-num}

\providecommand{\newblock}{}
\begin{thebibliography}{1}
\expandafter\ifx\csname url\endcsname\relax
  \def\url#1{{\tt #1}}\fi
\expandafter\ifx\csname urlprefix\endcsname\relax\def\urlprefix{URL }\fi
\providecommand{\eprint}[2][]{\url{#2}}

\bibitem{Asaka:2005an}
Asaka T, Blanchet S and Shaposhnikov M 2005 {\em Phys. Lett. B\/} {\bf 631}
  151--156 (\textit{Preprint} \eprint{hep-ph/0503065})

\bibitem{Dolgov:2000pj}
Dolgov A~D, Hansen S~H, Raffelt G and Semikoz D~V 2000 {\em Nucl. Phys. B\/}
  {\bf 580} 331--351 (\textit{Preprint} \eprint{hep-ph/0002223})

\bibitem{Dolgov:2000jw}
Dolgov A~D, Hansen S~H, Raffelt G and Semikoz D~V 2000 {\em Nucl. Phys. B\/}
  {\bf 590} 562--574 (\textit{Preprint} \eprint{hep-ph/0008138})

\bibitem{Ruchayskiy:2012si}
Ruchayskiy O and Ivashko A 2012 {\em JCAP\/} {\bf 10} 014 (\textit{Preprint}
  \eprint{1202.2841})

\bibitem{Sabti:2020yrt}
Sabti N, Magalich A and Filimonova A 2020 {\em JCAP\/} {\bf 11} 056
  (\textit{Preprint} \eprint{2006.07387})

\bibitem{Mastrototaro:2021wzl}
Mastrototaro L, Serpico P~D, Mirizzi A and Saviano N 2021 {\em Phys. Rev. D\/}
  {\bf 104} 016026 (\textit{Preprint} \eprint{2104.11752})

\bibitem{Hannestad:1995rs}
Hannestad S and Madsen J 1995 {\em Phys. Rev. D\/} {\bf 52} 1764--1769
  (\textit{Preprint} \eprint{astro-ph/9506015})

\bibitem{Pisanti:2007hk}
Pisanti O, Cirillo A, Esposito S, Iocco F, Mangano G, Miele G and Serpico P~D
  2008 {\em Comput. Phys. Commun.\/} {\bf 178} 956--971 (\textit{Preprint}
  \eprint{0705.0290})

\bibitem{Consiglio:2017pot}
Consiglio R, de~Salas P~F, Mangano G, Miele G, Pastor S and Pisanti O 2018 {\em
  Comput. Phys. Commun.\/} {\bf 233} 237--242 (\textit{Preprint}
  \eprint{1712.04378})

\end{thebibliography}

\end{document}